\documentclass[preprint,12pt]{elsarticle}

\usepackage{graphics}

\usepackage{amssymb}
\usepackage{bm}

\usepackage{amsthm}

\usepackage{xcolor}

\begin{document}

\begin{frontmatter}



\title{Majorana $\phi_0$-Junction in a Disordered Spin-orbit Coupling Nanowire with Tilted Magnetic Field}

\author[sysu]{Hong Huang}
\author[su]{Qi-Feng Liang}
\author[sysu]{Dao-Xin Yao \corref{cor2} }
\author[sysu]{Zhi Wang \corref{cor1}}
\cortext[cor1]{Corresponding author: Zhi Wang; Email: physicswangzhi@gmail.com; Tel: +86-20-84111107}
\cortext[cor2]{Corresponding author: Dao-Xin Yao; Email: yaodaox@mail.sysu.edu.cn; Tel: +86-20-84112078}

\address[sysu]{School of Physics, Sun Yat-sen University, Guangzhou 510275, China}
\address[su]{Department of Physics, Shaoxing University, Shaoxing 312000, China}

\begin{abstract}
 Majorana Josephson junctions in nanowire systems exhibit a pseudo-$4\pi$ period current-phase relation in the clean limit. In this work, we study how this current-phase relation responds to a tilted magnetic field in a disordered Majorana Josephson junction within the Bogoliubov-de Gennes approach. We show that the tilted magnetic field induces a $\phi_0$ phase shift to the current-phase relation. Most importantly, we find that this $\phi_0$-junction behavior is robust even in the presence of disorders.
\end{abstract}


\date{\today}
\end{frontmatter}

\section{Introduction}
Topological superconductors (TS) have attracted much attention recently because they host Majorana bound states (MBS)\cite{read,kitaevwire,aliceareview,beenakkerreview,franzreview}. MBSs are zero energy quarsi-particles which obey non-Abelian statistics. They can construct non-local qubits which are topologically protected from local electromagnetic de-coherence. The remarkable non-Abelian character of MBSs makes their braiding operations available for topological manipulations of the non-local qubits. Despite these braiding operations are insufficient for building universal gates, they are still recognized as a corner stone for the realistic quantum computation\cite{ivanov,teo,aliceanphy,Leijnse}.

MBSs are also useful for constructing superconducting flux qubits\cite{maier,huangwc}. The zero-energy feature of the MBSs makes them available for transporting {\it half} Cooper pairs. This process brings in a $4\pi$ period current-phase relation (CPR)\cite{golubov} in Majorana Josephson junctions, which is important for constructing new types of superconducting flux qubits.
The $4\pi$ periodicity of the CPR can be destroyed by quasiparticle poisoning processes\cite{wangsr}. This pseudo-$4\pi$ period CPR manifests as a skewed sine function. It serves as a signal for experimentally detecting MBSs in several systems.

Among all theoretically proposed systems, a promising candidate for TS is the nanowire superconductor hybrid structure\cite{lutchyn,sau2,oreg,stanescu}. The wire is subjected to strong spin-orbit coupling and Zeeman energy, which split the spin degenerate parabolic energy band into two sub-bands. If the chemical potential of the wire is fine tuned so that the Fermi surface intersect with only one of the sub-band, the spin-degree of freedom of the electrons near the Fermi surface is effectively eliminated. Then the proximity to a conventional superconductor could induce an effective spin-less p-wave superconducting gap on the wire which, according to the Kitaev model, will make the wire a topological superconductor. This type of hybrid systems was first theoretically proposed and then experimentally realized in InSb and GaSe nanowires in contact with conventional superconductors such as Nb and Al\cite{mourik,das,deng}. A bunch of signals for MBSs have been reported in these experiments. In particular, a robust zero energy peak was found in the differential conductance of a junction between a metal and the nanowire. Exponential protection of this zero energy peak is also found for a nanowire island\cite{albrecht}. These signals are consistent with the theoretical predictions based on the characters of the MBSs\cite{sau2,law,flensberg}. Even though there are unresolved issues in these experiments, it is more probable that MBSs are indeed successfully produced in these systems. Up to now, the search for MBSs is still an important task.

\begin{figure}[t]
\begin{centering}
\includegraphics[clip=true,width=0.5\textwidth]{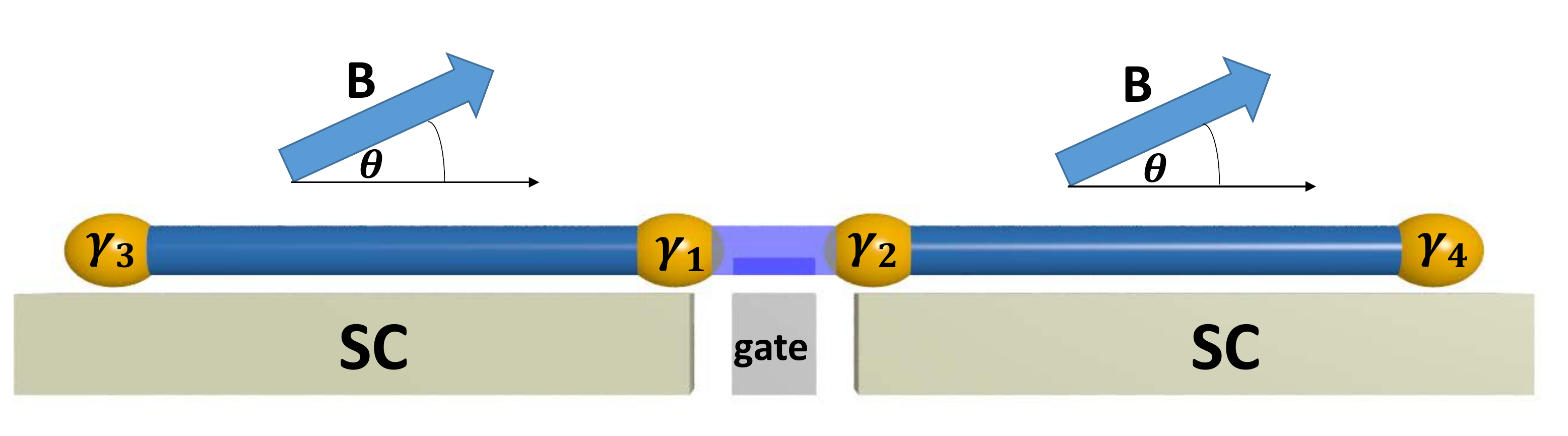}
\caption{(Color online). Schematic of a nanowire Majorana Josephson junction with a tilted Magnetic field. The nanowire is tuned into topological superconducting state. A voltage gate defines a tunneling barrier in the wire, and four MBSs appear nearby the junction and the two ends of the wire.}
\label{Fig1}
\end{centering}
\end{figure}

One defining feature of the topological superconductors is the topological Josephson effect\cite{beenakkerreview,lutchyn,van,law2,Asano}. In one-dimension topological Josephson junctions, the end MBSs carry a novel supercurrent with a $4\pi$ period current-phase relation (CPR).
In the static limit, the $4\pi$ periodicity may be destroyed by the quasiparticle poisoning by superconducting quasiparticles above the energy gap and other distant MBSs, which leaves a skewed $2\pi$ period CPR. Experimental search for the $4\pi$ period CPR in the dynamical processes and the skewed
$2\pi$ period CPR is still a hot topic\cite{oostinga,kurter,wiedenmann}.
The topological superconducting wire provides a perfect platform for studying topological Josephson effects. A voltage gate in the middle of the wire can easily create a potential barrier for electrons and create a Josephson junction. The height of the potential barrier is properly controlled by the applied voltage on the gate, which is able to tune the Josephson junction between the tunneling limit and ballistic limit. This nanowire junction has been build in experiments, and ac Josephson effect has been investigated with Shapiro step measurements\cite{jiang,san,rokhinson,badiane}. The missing of the first Shapiro steps was reported which is in agreement with the prediction of the $4\pi$ period CPR. Studying the Josephson effect on this nanowire junction may reveal different aspects of the topological Josephson effect.

One of the unconventional Josephson effect in topological nanowire junctions is the non-zero supercurrent under zero phase difference, the so-called $\phi_0$ junction behavior, in presence of a tilted magnetic field. This $\phi_0$ topological junction has been discussed in several systems, including a single wire junction and a ring shape junction\cite{wu,nesterov}. However, as far as we know, the effect of disorders has not been discussed. The tilting of the magnetic field in topological nanowire junctions has demonstrated rich phenomena\cite{alidoust,pientka,rex}. It is a natural question to consider the effect of a tilted magnetic field in the disordered nanowire Josephson junction.

In this work, we study the CPR of a topological nanowire Josephson junction in the presence of a tilted magnetic field and disorders. We use a realistic tight-binding model to describe the topological nanowire Josephson junction, and use Bogoliubov-de Gennes approach to obtain the current-phase relation. We show that the tilted magnetic field induces a Josephson current with no phase difference, making this nanowire junction a so-called $\phi_0$ Josephson junction\cite{buzdin,zazunov,yokoyama,huangzh,dolcini,campagnano,marra,szombati,huangzh2}. We systematically study different magnetic configurations, and show that the $\phi_0$ phase is proportional to the vertical component of the tilted magnetic field.  We then examine the influence of disorders. We consider both on-site and off-diagonal disorders, and find that the $\phi_0$ junction behavior is robust to the presence of moderate disorders. These results are relevant to the experimental search for Josephson signals of MBSs, and also might be useful for superconducting qubits based on Josephson junctions.

\section{Model and Numerical Method}
The system we consider, which is shown in figure \ref{Fig1}, is a topological nanowire Josephson junction. The wire requires a strong spin-orbit coupling, a proximity induced superconducting gap, an appropriate Zeeman energy and a well tuned chemical potential for entering the topological superconducting phase.
Let us first set up the minimal model Hamiltonian for a one-dimension nanowire. We adopt a tight-binding Hamiltonian for the wire,
\begin{eqnarray}
H_0 = - t \sum_{ \langle i , j \rangle, \sigma} c_{i , \sigma}^\dagger c_{j , \sigma} - \mu  \sum_i c^\dagger_{i,\sigma} c_{i,\sigma} ,
\label{H0}
\end{eqnarray}
where $t$ represents the hopping term and $\mu$ is the chemical potential. Here we only consider the nearest-neighbor hopping for simplicity. The spin orbit coupling is included through a Rashba term,
\begin{eqnarray}
H_{soc}=\eta  \sum_{i,  \sigma, \sigma'}^{n}  [c_{i+1,  \sigma}^{\dag}   (i\sigma_y)_{\sigma\sigma^\prime}   c_{i , \sigma'} +h.c.],
\label{Hsoc}
\end{eqnarray}
where $\eta$ represents the spin-orbit coupling strength. The superconducting paring gap in the nanowire is induced by proximity to a conventional s-wave superconductor. Therefore the effective s-wave superconducting gap is simply written as,
\begin{eqnarray}
H_{sc}(\phi)=\sum_{i=1}^{n}(\Delta^{i\phi}c_{i\uparrow}^\dag c_{i\downarrow}^\dag+h.c),
\label{Hsc}
\end{eqnarray}
where $\Delta$ is the amplitude of the superconducting gap and $\phi$ is the superconducting phase. The final piece for the topological superconductivity is the Zeeman energy. Here we want to discuss the effect of tilting magnetic field. Therefore, we consider contributions from both x and y components,
\begin{eqnarray}
H_{z}=\sum_{i, \sigma\sigma'}   c_{i, \sigma}^\dag (V_x \sigma_x+V_y \sigma_y)_{\sigma\sigma'}
c_{i, \sigma'} + h.c.,
\label{Hz}
\end{eqnarray}
where $V_x$ and $V_y$ represent the Zeeman energy from the magnetic field in the x and y directions.
Summing together, we arrive at the realistic Hamiltonian for the topological superconducting wire,
\begin{eqnarray}
H_w (\phi) = H_0 + H_{sc}(\phi)+ H_{soc} + H_z.
\label{Hw}
\end{eqnarray}
This wire enters the topological superconducting phase when the parameters satisfy the conditions $V_x^2 + V_y^2 > \mu^2 + \Delta^2$ and $V_y < \Delta$, which has been revealed in previous studies\cite{osca}.

We now consider the junction formed by two wires. We take the simplest tunneling junction which is formed by a potential barrier. The electron tunneling process through the potential barrier is described by a tunneling Hamiltonian,
\begin{eqnarray}
H_{T}=T_0  \sum _{\sigma} c_{L,\sigma}^{\dag} c_{R,\sigma}+h.c.,
\label{HT}
\end{eqnarray}
where $T_0$ represents the tunneling strength which is controlled by the potential barrier, $c_L$ and $c_R$ represent the electron annihilation operators on the ends of the two wires. Here we only consider the spin preserving tunneling process since a pure potential barrier does not flip the spin.
The Josephson junction consists of two wires which are connected with a tunneling Hamiltonian.
For simplicity, we consider two identical wires. The only difference between the two wires is the superconducting phase. Let us denote them as $\phi_L$ and $\phi_R$ for the left and the right wires, respectively. Their difference $\theta = \phi_L - \phi_R$ defines the phase difference across the junction. This phase difference drives a dc Josephson current $I_J (\theta)$. The total Hamiltonian of the system then is written as,
\begin{eqnarray}
H = H_w (\phi_L) + H_w (\phi_R) + H_T.
\label{H}
\end{eqnarray}

\begin{figure}[h]
\begin{centering}
\includegraphics[clip=true,width=0.8\columnwidth]{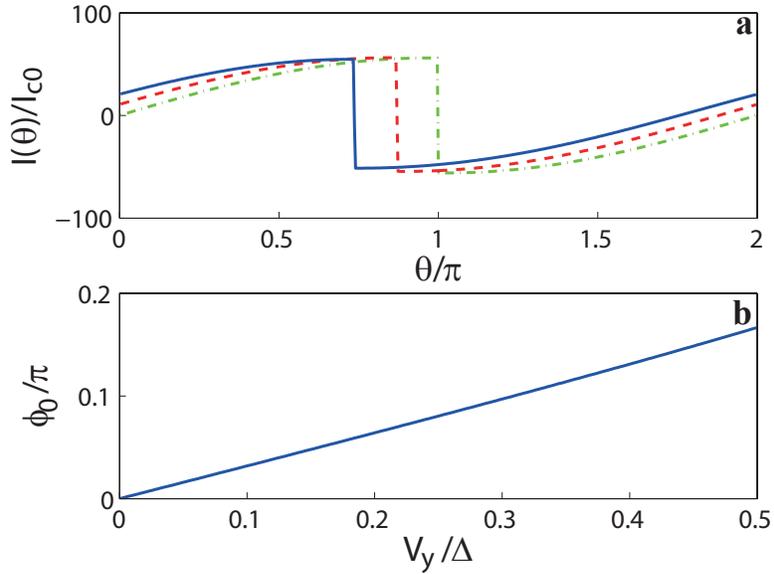}
\caption{(a) CPR of the junction with $V_y=0 $ (dash-dotted line), $V_y=0.009t$ (dashed line), and $V_y=0.018t$ (solid line). (b) The phase shift $\phi_0$ as a function of $V_y$. Parameters are taken as $V_x=0.09t$, $\mu= 0.18t$, $\Delta = 0.045t$, $\eta=0.122t$, $T_0=0.1t$, $I_{c0}=4\times10^{-4}{e\Delta}/{2\hbar}$.}
\label{Fig2}
\end{centering}
\end{figure}

Having established the tight-binding model for the system, we are now ready to investigate the Josephson current through the junction. We adopt the conventional Bogoliubov- de Gennes (BdG) method. In this approach, we introduce Bogoliubov operators
\begin{eqnarray}
c_{i\uparrow}=\sum_{n}\mu_{n\uparrow}\gamma_n+\nu_{n\uparrow}^\ast\gamma_n^\dag\\
c_{i\downarrow}=\sum_{n}\mu_{n\downarrow}\gamma_n+\nu_{n\downarrow}^\ast\gamma_n^\dag,
\end{eqnarray}
\label{BdG}
where $\mu$ and $\nu$ are transformation coefficients. We numerically diagonalize the entire Hamiltonian with this transformation, $H=\sum_{n}E_n\gamma_{n}^\dag\gamma_{n}$, where we obtain the energy spectrum $E_n$. We smoothly change the phase difference $\theta$ across the junction. The energy spectrum changes accordingly. We then get a phase difference dependence of the  energy spectrum $E_n (\theta)$. The derivative of the phase difference on this energy spectrum gives the CPR of the junction with a formula of $I(\theta)= \sum_ { E_n < 0 } \frac{e} {\hbar} \frac{d}{d \theta} E_n(\theta)$. In our numerical approach, we first solve an eigen-problem to obtain the energy spectrum $E_n (\theta)$, and then using the above formula to calculate the CPR of the junction.

\section{Results and Discussions}
\subsection{The $\phi_0$ phase shift induced by tilting magnetic field}
The Hamiltonian showed in equation \ref{H} has been well studied when the Zeeman field is in parallel to the nanowire with zero vertical component $V_y = 0$. It is found that zero energy MBSs appear when the Zeeman field exceeds a critical value $V_x^2 > \mu^2 + \Delta^2$. Disorders may complicate the problem but does not affect the qualitative conclusion. A topological quantum phase transition always exists around $V_x^2 = \mu^2 + \Delta^2$. This phase transition connects the topological phase with MBSs and the trivial phase without MBSs. The Josephson effect is also studied when a voltage potential is implemented in the middle of the wire. It is found that a strong enhancement of the critical current and a pseudo-$4\pi$ period current-relation exists in the topological regime\cite{huangh}.

Now let us investigate the tilted Zeeman field with a finite vertical component $V_y$. We consider the topological regime in which the parallel Zeeman field is taken to be large $V_y \ll V_x$. The vertical component $V_y$ is small, therefore should not alter the topology of the superconducting nanowire. However, they could make quantitative contributions to the Josephson currents. We numerically study the CPR of the topological Josephson junction with different values of $V_y$, and show the results of three typical values in figure \ref{Fig2}a. For parallel Zeeman field with zero vertical component $V_y = 0$, we see that the CPR is almost an exact function of $I(\theta) = \pm I_c  \sin \frac{\theta}{2}$, where the sign changes from plus to minus at $\theta = \pi$. This is exactly the skewed pseudo-$4\pi$ period CPR from the MBS channel. We note that the jump of the CPR comes from the coupling between MBSs which is different from the conventional one induced by ballistic transport.
When the Zeeman field is tilted from the parallel direction and the vertical component arises, we find that the shewed shape of the CPR remains the same. This indicates that the Josephson current is still mainly carried by the MBS channel. However, a $\phi_0$ phase shift appears in the CPR, which is explicitly represented by a non-zero Josephson current at zero phase difference. This is the so-called $\phi_0$ Josephson junction. When the vertical component of the tilted Zeeman field is larger, we find that the phase shift $\phi_0$ also becomes larger. Since the vertical component only slightly change the critical current, we have a larger Josephson current for zero phase difference. This $\phi_0$ pseudo-$4\pi$ CPR is unconventional, and would be potentially useful for identifying the MBSs.

For a more general study, we calculate the dependence of the phase shift on the vertical component of the tilted Zeeman field. In figure \ref{Fig2}b, we show the phase shift $\phi_0$ as a function of the Zeeman energy $V_y$. We find a linear function which is the simplest form we would expect, contrasting to the situation in trivial topological phase\cite{wu}. The simple relation between $V_y$ and $\phi_0$ imply a tunable 
Josephson $\phi_0$ current.
\begin{figure}[h]
\begin{centering}
\includegraphics[clip=true,width=0.8\columnwidth]{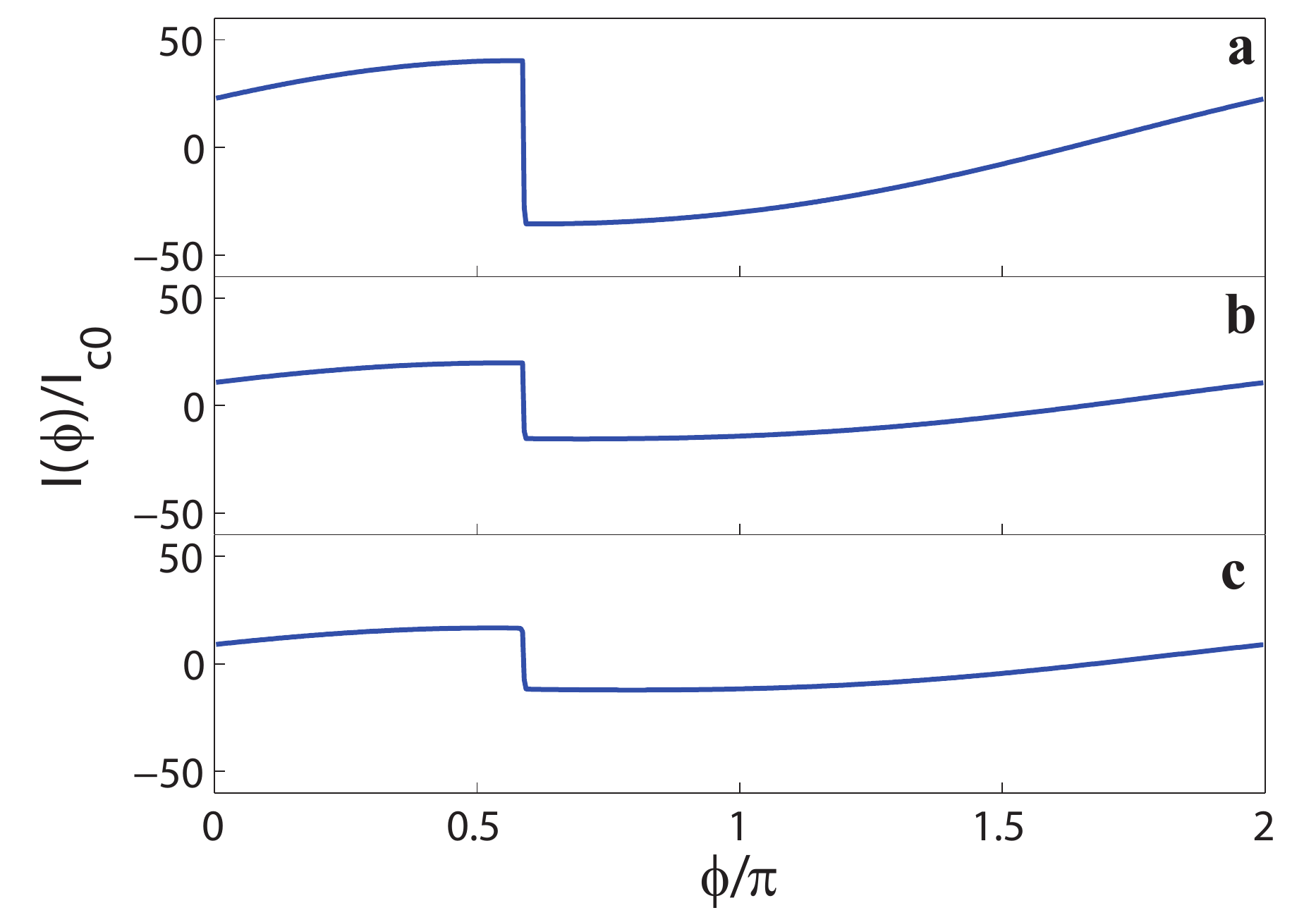}
\caption{CPR of the junction under a 10$\%$ random distribution of on-site impurities with different potential of (a) $V_i = 0.09 t$, (b) $V_i = 0.18 t$,  and (c) $V_i = 0.27 t$. Other parameters are taken the same as in figure 2 with $V_y=0.027 t$.}
\label{Fig3}
\end{centering}
\end{figure}

\begin{figure}[h]
\begin{centering}
\includegraphics[clip=true,width=0.8\columnwidth]{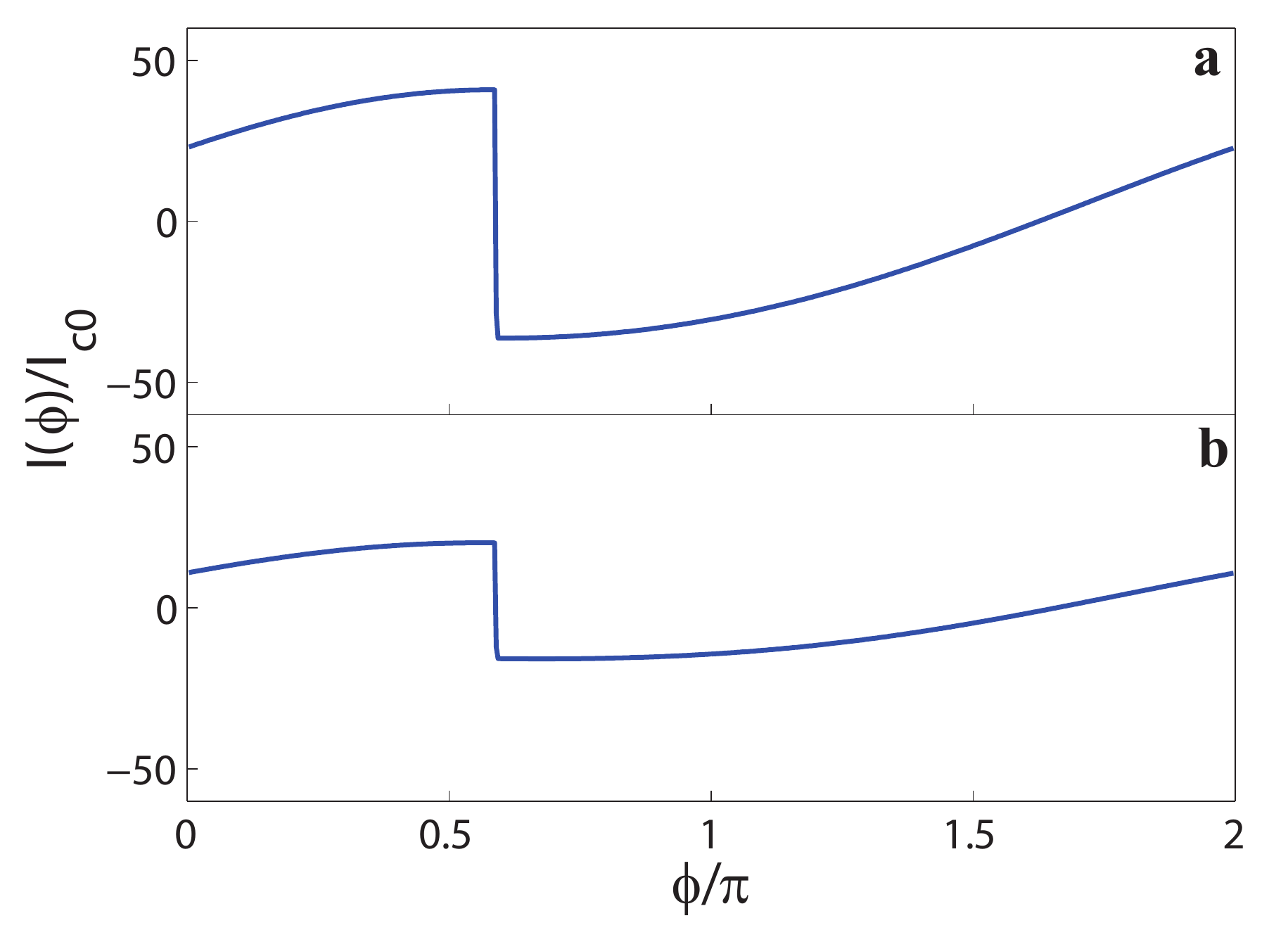}
\caption{(Color online) CPR of the junction under an impurity potential of $V_i = 0.09 t$ with a random distribution of (a) $20\%$ and (b) $50\%$. Other parameters are taken the same as in figure \ref{Fig3}.}
\label{Fig4}
\end{centering}
\end{figure}

\subsection{The Disordered topological Josephson junctions}
 In the study of topological nanowire systems, disorders must be considered within any theoretical analysis because they are important obstacles in any one-dimension structure.
The effect of disorder has been studied in odd frequency topological junctions, and universal transport behavior irrespective of disorders has been revealed\cite{Asano}. In one-dimension topological superconductors, disorders may reduce the superconducting gap, therefore eliminate the conductance signal for gap closure at the topological quantum phase transition point\cite{mourik}. 
With these considerations, it is a natural task to investigate the stability of the $\phi_0$ junction behavior of the topological Josephson junction in the presence of disorders. For this purpose, we add two different types of disorders into the nanowire, and calculate the CPR with varies of disorder concentrations.

We first consider the simplest disorders induced by unitary impurities.
Each impurity modulates the local chemical potential, which can be described by an on-site Hamiltonian of the form,
\begin{eqnarray}
H_i = V_i \sum_{\sigma}  c^\dagger_{\sigma} c_{\sigma},
\label{Hion}
\end{eqnarray}
where $V_i$ is the local potential from the impurity. For simplicity, we consider identical impurities therefore the local potentials are the same for all impurity sites. The impurities are randomly distributed in the nanowire with the concentration varying from small to large. This Hamiltonian for the disorders is the simplest one which accounts for the doping of the same type of unitary impurities. However, it already provides qualitative features for the CPR. Let us first look at a fixed percentage of disorders with different local impurity potentials. We demonstrate the CPR for three typical cases with small and large impurity potentials in figure 3. When the impurity potential is relatively small with $V_i = 0.09 t$, we see that the CPR is almost the same as the one without impurities. This implies that weak disorders do not influence the CPR of the topological junction. While the impurity potential is doubled to $V_i = 0.18 t$ and even tripled to $V_i=0.27 t$, the $\phi_0$ phase shift of the CPR are still preserved.
These results demonstrate that the $\phi_0$ junction behavior is robust in the presence of moderate disorders.

To check the generality of these disordered results, we also consider the case that the impurity potential is fixed, while its concentration is increased. In figure 4, we show the results of two different impurity concentrations with the same impurity potential $V_i = 0.09 t$. We find that both the skewed shape of the CPR and the $\phi_0$-junction behavior are robust even in the presence of a large concentration of impurities, though the value is suppressed significantly by the large concentration of impurities.

\begin{figure}[h]
\begin{centering}
\includegraphics[clip=true,width=0.8\columnwidth]{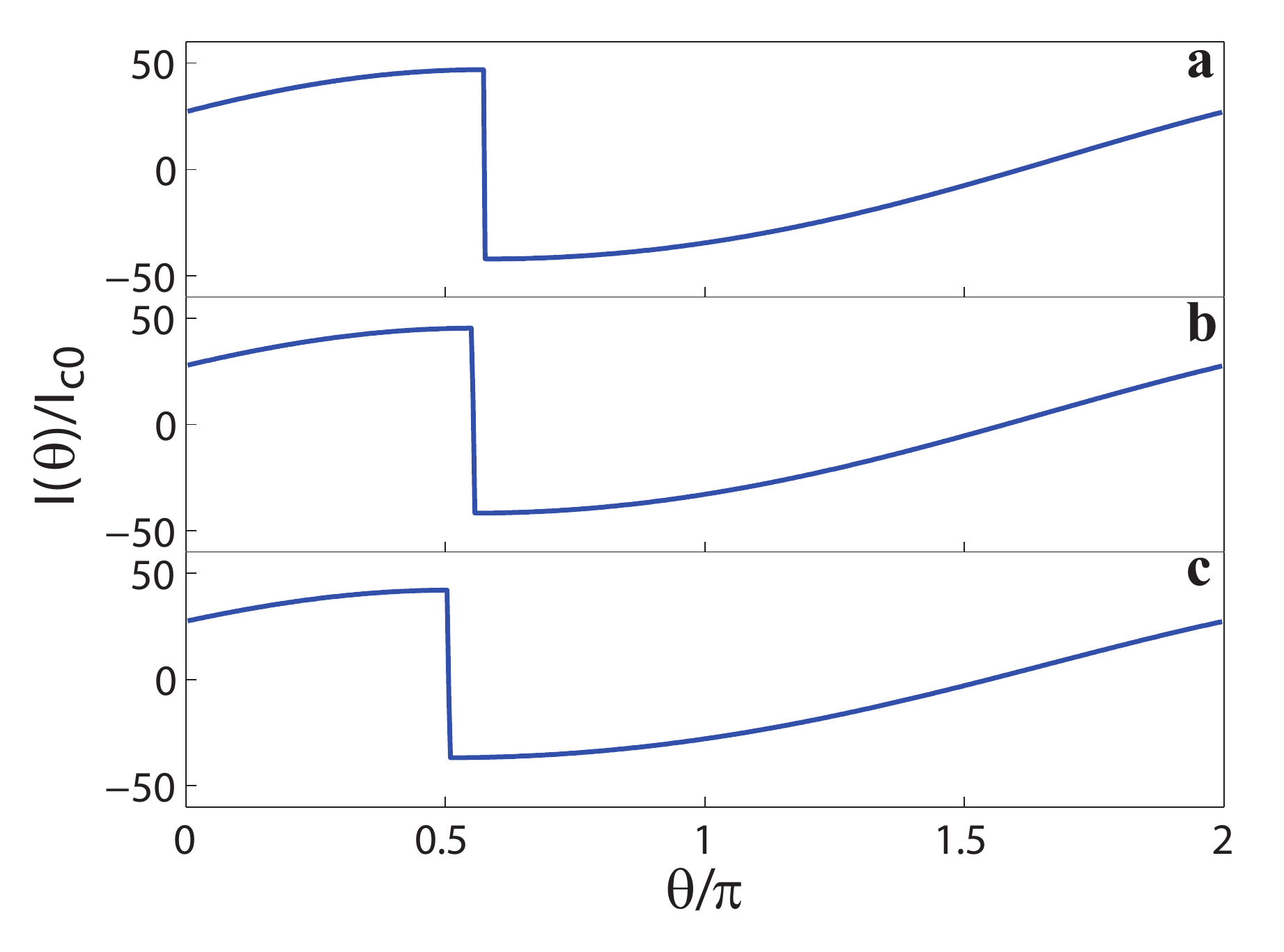}
\caption{CPR of the junction under a 10$\%$ random distribution of off-diagonal disorders with the magnitude of (a) $ V_o = 0.2 \Delta$, (b) $ V_o = 0.6 \Delta$, and (c) $V_o =  \Delta$. Other parameters are taken the same as in figure \ref{Fig3}.}
\label{Fig5}
\end{centering}
\end{figure}

\begin{figure}[h]
\begin{centering}
\includegraphics[clip=true,width=0.8\columnwidth]{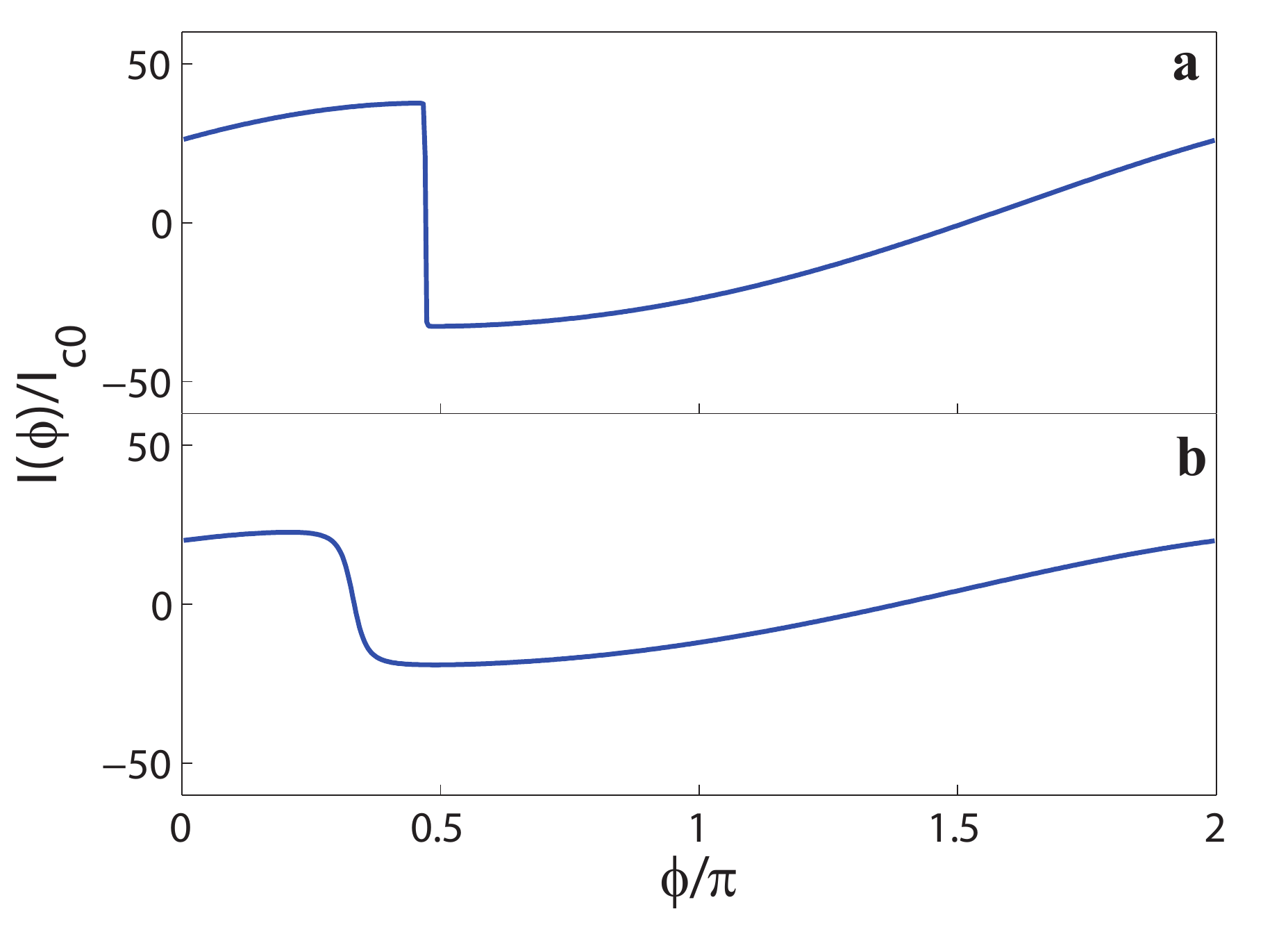}
\caption{(Color online) CPR of the junction under a disorder potential of $V_o =0.6 \Delta$ with a random distribution of (a) $30\%$ and (b) $50\%$ of disorders. Other parameters are taken the same as in figure \ref{Fig3}.}
\label{Fig6}
\end{centering}
\end{figure}

Next we consider the off-diagonal disorders, which come from the suppression of the local superconducting paring gap by the impurities. We note that this off-diagonal effect should be strong in the topological nanowire junction, since the proximity induced superconductivity is effectively p-wave for achieving topological superconducting phase. As before, we consider the simplest type of disorder Hamiltonian,
\begin{eqnarray}
H_i= - V_{o} c_{\uparrow}^\dag c_{\downarrow}^\dag+h.c,
\label{Hioff}
\end{eqnarray}
where $V_o$ represents the local suppression of the pairing function. We note the $V_o$ is smaller than $\Delta$ since the superconducting paring gap is at most suppressed to zero. In figure 5, we show the results for ten percent of off-diagonal disorders with three typical disorder potentials. We see that the off-diagonal disorders have a much smaller effect on both the skewed shape of the CPR and the $\phi_0$ junction behavior. These two features are nearly intact even for the large possible disorder strength of $V_o = \Delta$, which means that the superconducting gap is fully suppressed by the impurity.
 We also present the results of increasing disorder concentrations in figure 6, while the disorder strength is fixed to $V_o = 0.6 \Delta$. We find that the skewed shape of the CPR is slightly modulated for large impurity concentrations. However, the $\phi_0$ junction behavior is not influenced.

The disorders may have diagonal and off-diagonal contributions simultaneously. The qualitative results would be the same for these more general conditions since we have shown that the diagonal and off-diagonal disorders exhibit similar results.
With these numerical simulations, we conclude that the $\phi_0$ junction behavior induced by the tilting Zeeman field is robust in the presence of disorders. Therefore, the $\phi_0$ junction behavior might be helpful for the detection of MBSs in experiments, in which the disorders are unavoidable.

\section{Conclusion}
In summary, we have used the Bogoliubov-de Gennes approach to study the Josephson current through a topological nanowire junction. In particular, we study the effect of a tilted magnetic field with and without disorders. We find that the tilted magnetic field induces a phase shift of the current-phase relation of the junction, making it become a $\phi_0$ Josephson junction. This $\phi_0$-junction behavior is robust in the presence of both diagonal and off-diagonal disorders. Therefore, it might be helpful in detecting Majorana bound states. In the meantime, the $\phi_0$ junction is also useful for building quantum devices. 

\section*{Acknowledgements}
This work is supported by NSFC-11304400, NSFC-61471401, and SRFDP-20130171120015. Q.F.L is supported by NSFC-11574215. D.X.Y. is supported by NSFC-11074310, NSFC-11275279, SRFDP-20110171110026, and NBRPC-2012CB821400.

\section*{References}




\bibliographystyle{elsarticle-num}
\bibliography{<your-bib-database>}



\end{document}